\documentclass[onecolumn]{aastex631}

\usepackage{listings}
\usepackage{amsmath}
\usepackage{multirow}

\shorttitle{The Astronomy Commons Platform}
\shortauthors{Stetzler et al.}

\graphicspath{{./}{figures/}}

\begin{document}

\title{The Astronomy Commons Platform: A Deployable Cloud-Based Analysis Platform for Astronomy}

\correspondingauthor{Steven Stetzler}
\email{stevengs@uw.edu}

\author[0000-0002-7712-6678]{Steven Stetzler}
\affiliation{DiRAC Institute and the Department of Astronomy, University of Washington, Seattle, USA}

\author[0000-0003-1996-9252]{Mario Juri\'c}
\affiliation{DiRAC Institute and the Department of Astronomy, University of Washington, Seattle, USA}

\author[0000-0002-5828-6211]{Kyle Boone}
\affiliation{DiRAC Institute and the Department of Astronomy, University of Washington, Seattle, USA}

\author[0000-0001-5576-8189]{Andrew Connolly}
\affiliation{DiRAC Institute and the Department of Astronomy, University of Washington, Seattle, USA}

\author[0000-0002-0558-0521]{Colin T. Slater}
\affiliation{DiRAC Institute and the Department of Astronomy, University of Washington, Seattle, USA}

\author[0000-0002-2651-243X]{Petar Ze\v{c}evi\'c}
\affiliation{Faculty of Electrical Engineering and Computing, University of Zagreb, Croatia}
\affiliation{Visiting Fellow, DIRAC Institute, University of Washington, Seattle, USA}

\begin{abstract}

We present a scalable, cloud-based science platform solution designed to enable next-to-the-data analyses of terabyte-scale astronomical tabular datasets. The presented platform is built on Amazon Web Services (over Kubernetes and S3 abstraction layers), utilizes Apache Spark and the Astronomy eXtensions for Spark for parallel data analysis and manipulation, and provides the familiar JupyterHub web-accessible front-end for user access. We outline the architecture of the analysis platform, provide implementation details, rationale for (and against) technology choices, verify scalability through strong and weak scaling tests, and demonstrate usability through an example science analysis of data from the Zwicky Transient Facility's 1Bn+ light-curve catalog. Furthermore, we show how this system enables an end-user to iteratively build analyses (in Python) that transparently scale processing with no need for end-user interaction.

The system is designed to be deployable by astronomers with moderate cloud engineering knowledge, or (ideally) IT groups. Over the past three years, it has been utilized to build science platforms for the DiRAC Institute, the ZTF partnership, the LSST Solar System Science Collaboration, the LSST Interdisciplinary Network for Collaboration and Computing, as well as for numerous short-term events (with over 100 simultaneous users). A live demo instance, the deployment scripts, source code, and cost calculators are accessible at \url{http://hub.astronomycommons.org/}. 

\end{abstract}


\section{Introduction} \label{sec:intro}

Today's astronomy is undergoing a major change. Historically a data-starved science, it is being rapidly transformed by the advent of large, automated, digital sky surveys into a field where terabyte and petabyte data sets are routinely collected and made available to researchers across the globe.

The Zwicky Transient Facility \citep[ZTF;][]{2019PASP..131a8002B, 2019PASP..131g8001G, 2020PASP..132c8001D, 2019PASP..131a8003M} has engaged in a three-year mission to monitor the Northern sky. With a large camera mounted on the Samuel Oschin 48-inch Schmidt telescope at Palomar Observatory, the ZTF is able to monitor the entire visible sky almost twice a night. Generating about 30~GB of nightly imaging, ZTF detects up to 1,000,000 variable, transient, or moving sources (or alerts) every night, and makes them available to the astronomical community \citep{Patterson_2018}. Towards the middle of 2024, a new survey, the Legacy Survey of Space and Time  \citep[LSST;][]{lsstOverview}, will start operations on the NSF Vera C. Rubin Observatory. Rubin Observatory's telescope has a mirror almost seven times larger than that of the ZTF, which will enable it to search for fainter and more distant sources. Situated in northern Chile, the LSST will survey the southern sky taking ${\sim}1,000$ images per night with a 3.2 billion-pixel camera with a ${\sim}10$ deg$^2$ field of view. The stream of imaging data (${\sim}6$PB/yr) collected by the LSST will yield repeated measurements (${\sim}100$/yr) of over 37 billion objects, for a total of over 30 trillion measurements by the end of the next decade. These are just two examples, with many others at similar scale either in progress \citep[Kepler, Pan-STARRS, DES, GAIA, ATLAS, ASAS-SN;][]{2010SPIE.7733E..0EK, 2016MNRAS.460.1270D, 2016A&A...595A...1G,2018PASP..130f4505T,2014AAS...22323603S} or planned \citep[Roman, Euclid;][]{2015arXiv150303757S,2014IAUS..306..375S}. They are being complemented by numerous smaller projects ($\lesssim$\$1M scale), contributing billions of more specialized measurements.

This 10-100x increase in survey data output has not been followed by commensurate improvements in tools and platforms available to astronomers to manage and analyze those catalogs. Most survey-based studies today are performed by navigating to archive websites, entering (very selective) filtering criteria to download ``small'' (${\sim}10$s of millions of rows; ${\sim}10$GB) subsets of catalog products. Those subsets are then stored locally and analyzed using custom routines written in high-level languages (e.g., Python or IDL), with the algorithms generally assuming in-memory operation. With the increase in data volumes and subsets of interest growing towards the ${\sim}100$GB-1TB range, this mode of analysis is becoming infeasible.

One solution is to provide astronomers with access to the data through web portals and \textit{science platforms} -- rich gateways exposing server-side code editing, management, execution and result visualization capabilities -- usually implemented as notebooks such as Jupyter \citep{jupyter-notebooks} or Zeppelin \citep{zepplin}. These systems are said to {\em bring the code to the data}, by enabling computation on computational resources co-located with the data and providing built-in tools to ease the process of analysis. For example, the Rubin Observatory/LSST has designed \citep{LSE-319, LDM-542} and implemented  a science platform suitable for accessing and visualizing data from the LSST, with deployments hosted on both on-premises hardware and Google Cloud \citep{rsp_google}.\footnote{See as well \url{https://data.lsst.cloud/}} While such science platforms are a major step forward in working with large datasets, they still have some limitations when deployed on on-premises hardware or traditional HPC systems. These systems can suffer from having insufficient computing next to the data: all users of shared HPC resources are familiar with ``waiting in the queue'' due to over subscription. Science platforms built on cloud computing resources will find it much easier to provide computing resources according to user demand: this is the promise of ``elastic'' computing in the cloud.

Secondly, even when surveys deploy distributed SQL databases for serving user queries \citep[e.g. Qserv in the case of LSST;][]{qserve}, user analysis is still not easily parallelized -- query requests and results are bottlenecked at one access point which severely limits scalability. In contrast, the system we describe and implement provides direct, distributed access to data for a user’s analysis code. Finally, current science platforms do not tackle the issue of working on multiple large datasets at the same time -- if they're in different archives, they still have to be staged to the same place before work can be done. In other words, they continue to suffer from availability of computing, being I/O-bound, and geographic dislocation.

We therefore need to not only bring the code to the data, but also {\em bring the data together}, co-locate it next to an (ideally limitless) reservoir of computing capacity, with I/O capabilities that can scale accordingly. Furthermore, we need to make this system {\em usable}, by providing astronomer-friendly frameworks for working with extremely large datasets in a scalable fashion. Finally, we need to provide a user-interface which is accessible and familiar, with a shallow learning curve.

We address the first of these challenges by utilizing the Cloud to supply data storage capacity, effective dataset co-location, I/O bandwidth, and (elastic) compute capability. This work utilizes the Amazon Web Services (AWS) cloud, leveraging Amazon Simple Storage Service (Amazon S3) for storage and access to TB+ sized tabular data sets (catalogs) and Amazon Elastic Cloud Compute (Amazon EC2) for elastic computing. \cite{aws_gateways} have investigated using the same services for scalable storage, access, and processing of image data. The second challenge is addressed by extending the Astronomy eXtensions for Spark \citep[AXS;][]{zecevic}, a distributed database and map-reduce like workflow system built on the industry-standard Apache Spark \citep{spark} engine, to work in this cloud environment. Spark allows the execution of everything from simple ANSI SQL-2011 compliant queries to complex distributed workflows, all driven from Python. When using Spark, data can be sourced from a number of storage solutions and a variety of formats, including FITS \citep{spark_fits}. Finally, a JupyterHub facade provides a user-friendly entry-point to the system. Additionally, we make it possible for IT groups (or advanced users) to easily deploy this entire system for use within their departments, as an out-of-the-box solution for cloud-based astronomical data analysis.

The combination of these technologies allows the researcher to migrate ``classic'' subset-download-analyze workflows with little to no learning curve, while providing an upgrade path towards large-scale analysis. We validate the approach by deploying a cloud-based platform for accessing and analyzing a 1 billion+ light-curve catalog from the ZTF (a precursor to LSST), and demonstrate it can be successfully used for exploratory science.

\section{A Platform for User-Friendly Scalable Analysis of Large Astronomical Datasets} \label{sec:platform}

We begin by introducing the properties of cloud systems that make them especially suitable for scalable astronomical analysis platforms, discuss the overall architecture of our platform, its individual components, and performance.

\subsection{The Cloud} \label{sec:cloud}

Traditionally, computing infrastructure was acquired and maintained close to the group utilizing the resource. For example, a group led by a faculty member would purchase and set up one or more machines for a particular problem, or (on a larger scale) a university may centralize computing resources into a common cluster, shared with the larger campus community. These acquisitions -- so-called ``on-premise'' computing -- are capital heavy (require a large initial investment), require local IT knowledge, and allow for a limited variety of the systems being purchased (e.g., a generic Linux machine for a small group, or standardized types of nodes for an HPC cluster).

Cloud services move this infrastructure (and the work to maintain it) away from the user, and centralize it with the cloud provider. The infrastructure is provided as a service: individual machines, entire HPC clusters, as well as higher-order services (databases, filesystes, etc.) are {\em rented} for the time the resource is needed, rather than purchased.

They are billed proportional to usage; virtual machines are typically rented by the second, virtual networks priced by bandwidth usage, and virtual storage priced by storage size per unit time. These components are provisioned by the user on-demand, and are built to be ``elastic.'' One can typically rent several hundred virtual machines and provision terabytes of storage space with an expectation that it will be delivered within minutes and then release these resource back to the cloud provider at will. This usage and pricing model offers the unique benefit of providing access to affordable computing at scale. One can rent hundreds of virtual machines for a short period of time (just the execution time of a science workflow) without investing in the long-term support of the underlying infrastructure. In addition, cloud providers typically offer managed storage solutions to support reading/writing data to/from all of these machines. These so-called ``object stores'' are highly available, highly durable, and highly scalable stores of arbitrarily large data volumes. For example, Amazon Simple Storage Solution (Amazon S3) provides scalable, simultaneous access to data through an Application Programming Interface (API) over a network.\footnote{Amazon S3 uses a REST API with HTTP.} S3 supports very high throughput at the terabit-per-second level assuming storage access patterns are optimized.\footnote{This is detailed in the S3 documentation: \url{https://docs.aws.amazon.com/AmazonS3/latest/dev/optimizing-performance.html}} Once a solution for scalable storage is added to the mix, cloud computing systems start to resemble the traditional supercomputers many scientists are already familiar with for running simulations and performing large-scale data analysis.

\subsection{Orchestrating cloud applications: Kubernetes}

The pain point that remains in managing and developing applications for the cloud is the problem of orchestration: it can become burdensome to write custom software for provisioning and managing cloud resources, and there is a danger of cloud ``lock-in'' occurring when software applications become too strongly coupled with the cloud provider's API. The open source community has developed orchestration tools, like Kubernetes, to address this issue.\footnote{The Kubernetes documentation provides a thorough and beginner-friendly introduction to the software: \url{https://kubernetes.io/docs/}}

\begin{figure}
    \centering
    \includegraphics[width=\linewidth]{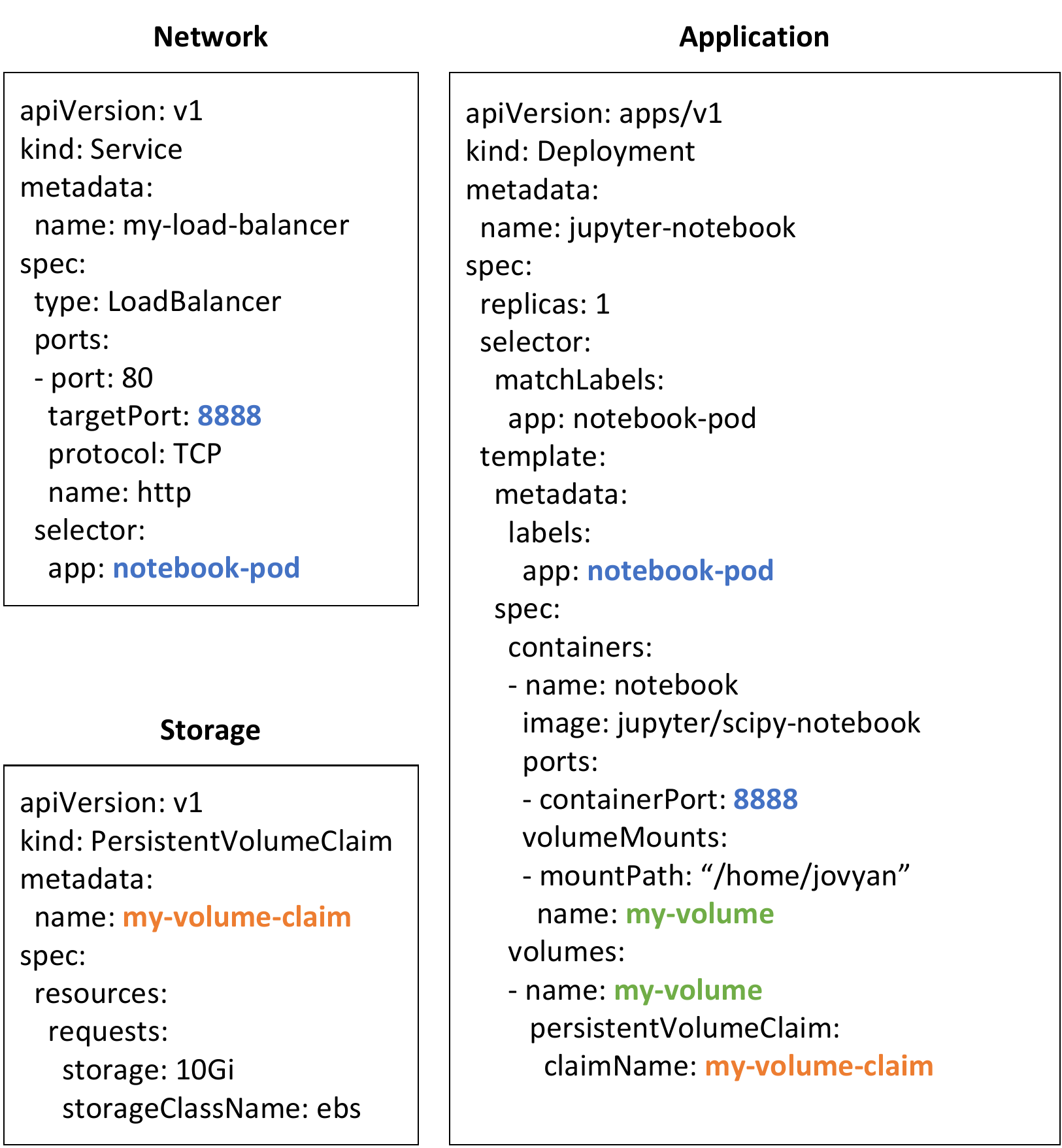}
    \caption{An illustration of the structure and composition of YAML-formatted text specifying Kubernetes objects that together create a functional and internet-accessible Jupyter notebook server. The Jupyter notebook application is created as a Pod on the cluster (right). Networking objects (top left) specify how a public-facing load balancer can be connected to the Jupyter notebook Pod (\texttt{notebook-pod}) on a certain port (\texttt{8888}). Storage objects trigger the creation of, for example, hard drive disk space from the cloud provider (bottom left). Colored text indicate how the files are linked to support one another: blue indicates how network and application are linked, orange how application and storage are linked, and green how storage volumes are mounted into the filesystem of the application.}
    \label{fig:k8s_objects}
\end{figure}

Kubernetes is used to schedule software applications packaged in Docker images and run as Docker containers on a cluster of computers (physical or virtual machines) while handling requests for and the provisioning of cloud resources to support running those containers.\footnote{Docker isolates software programs at the level of the operating system, in contrast to virtual machines which isolate operating systems from one another at the hardware level. See \url{https://www.docker.com/} and \url{https://docs.docker.com/} for more information.} Kubernetes provides a cloud-agnostic API, accessible over over a network using HTTP(S), to describe cloud resources as ``representational state transfer'' (REST) objects. For example, cluster storage is described using ``Persistent Volume'' objects, requests for that storage use ``Persistent Volume Claim'' objects, and networking utilities like routing, port-forwarding, and load balancing use ``Service'' objects. An application that runs using one or more containers is specified using a ``Pod'' object. If the application requires it, the Pod object can reference storage objects and service objects by name to link an application to these resources. In addition, the Pod object allows one to impose CPU and memory limits on an application or assign the application to a certain node, among other features. Each Kubernetes object is described using YAML, a human-readable format for storing configuration information (lists and dictionaries of strings and numbers).\footnote{See \url{https://yaml.org/} for specification and implementations.} Figure~\ref{fig:k8s_objects} shows an example set of YAML-formatted text describing Kubernetes objects that together would link a Jupyter notebook server backed by a 10 GiB storage device to an internet-accessible URL.

The Kubernetes core service, a set of software called the control plane, is responsible for maintaining an API server accessible within (and potential externally to) the cluster, maintaining a database of the objects created so far, assigning pods (applications) to nodes in the cluster in a way that respects their constraints, keeping track of the general state of the cluster, and handling aspects of networking within the cluster and through the cloud provider. Additional components of the Kubernetes core code (or third-party plugins) handle provisioning of virtual hardware from the cloud provider to satisfy requirements that cannot be met by current cluster resources. As an example on AWS, an outstanding request for a Service requiring a load balancer will be fulfilled by creating an AWS Elastic Load Balancer (ELB) or Application Load Balancer (ALB). Similarly, an outstanding request for a Persistent Volume will be fulfilled by creating an Amazon Elastic Block Store (EBS) volume. Finally, applications can be scheduled on the cluster that modify the cluster state. In particular, the Kubernetes Cluster Autoscaler interacts with the cloud provider to terminate underutilized nodes or add new nodes when there are pods that cannot be scheduled given the current number of nodes.\footnote{\url{https://github.com/kubernetes/autoscaler}} The handling of hardware provisioning from the cloud provider by administrative software in the Kubernetes control plane, through Kubernetes plugins, and through applications running in the cluster allows additional user applications to remain decoupled from the cloud provider's API.

\subsection{System Architecture}
\label{sec:arch}

Cloud systems offer unique infrastructure elements that help support a system for scalable science analysis. Virtual machines can be rented in the hundreds or thousands to support large computations, each accessing data in a scalable manner from a managed service. Orchestration layers, like Kubernetes, ease the process of running science software on cloud resources. In this section, we discuss how we leverage cloud infrastructure to build such a platform. Underlying this platform are four key components: 
\begin{enumerate}
    \item An interface for computing. We use the Jupyter ecosystem: a JupyterHub deployment based on the \texttt{zero-to-jupyterhub} project that creates Jupyter notebook servers on our computing infrastructure for authenticated users. A Jupyter notebook server provides a web-interface to interactively run code on a remote machine alongside a set of pre-installed software libraries.\footnote{See \url{https://zero-to-jupyterhub.readthedocs.io/} and \url{https://github.com/jupyterhub/zero-to-jupyterhub-k8s}.}
    \item A scalable analytics engine. We use Apache Spark, an industry standard tool for distributed data querying and analysis, and the Astronomy eXtensions to Spark (AXS).
    \item A scalable storage solution. We use Amazon Simple Storage Solution (S3). Amazon S3 is a managed object store that can store arbitrarily large data volumes and scale to an arbitrarily large number of requests for this data.
    \item A deployment solution. We've developed a set of Helm charts and bash scripts automating the deployment of this system onto the AWS cloud.\footnote{For Helm, see \url{https://helm.sh/}.}
\end{enumerate}

Each of these components are largely disconnected from one another and can be mixed and matched with other drop-in solutions.\footnote{Zepplin notebooks, among other tools, compete with Jupyter notebooks for accessing remote computers for analysis and data visualization. Dask is a competing drop-in for Apache Spark that scales Python code natively. A Lustre file system could be a drop-in for Amazon S3. Amazon EFS, a managed and scalable network filesystem, is also an option. Kustomize is an alternative to Helm.} Aside from the deployment solution, each of these components are comprised of simple processes communicating with each other through an API over a network. This means that each solution for (1), (2), and (3) is largely agnostic to the choice of running on a bare-metal machine, inside a virtual machine (VM), inside a Linux container, or using a managed cloud service as long as each component is properly networked.

\begin{figure}
\centerline{\includegraphics[width=\linewidth]{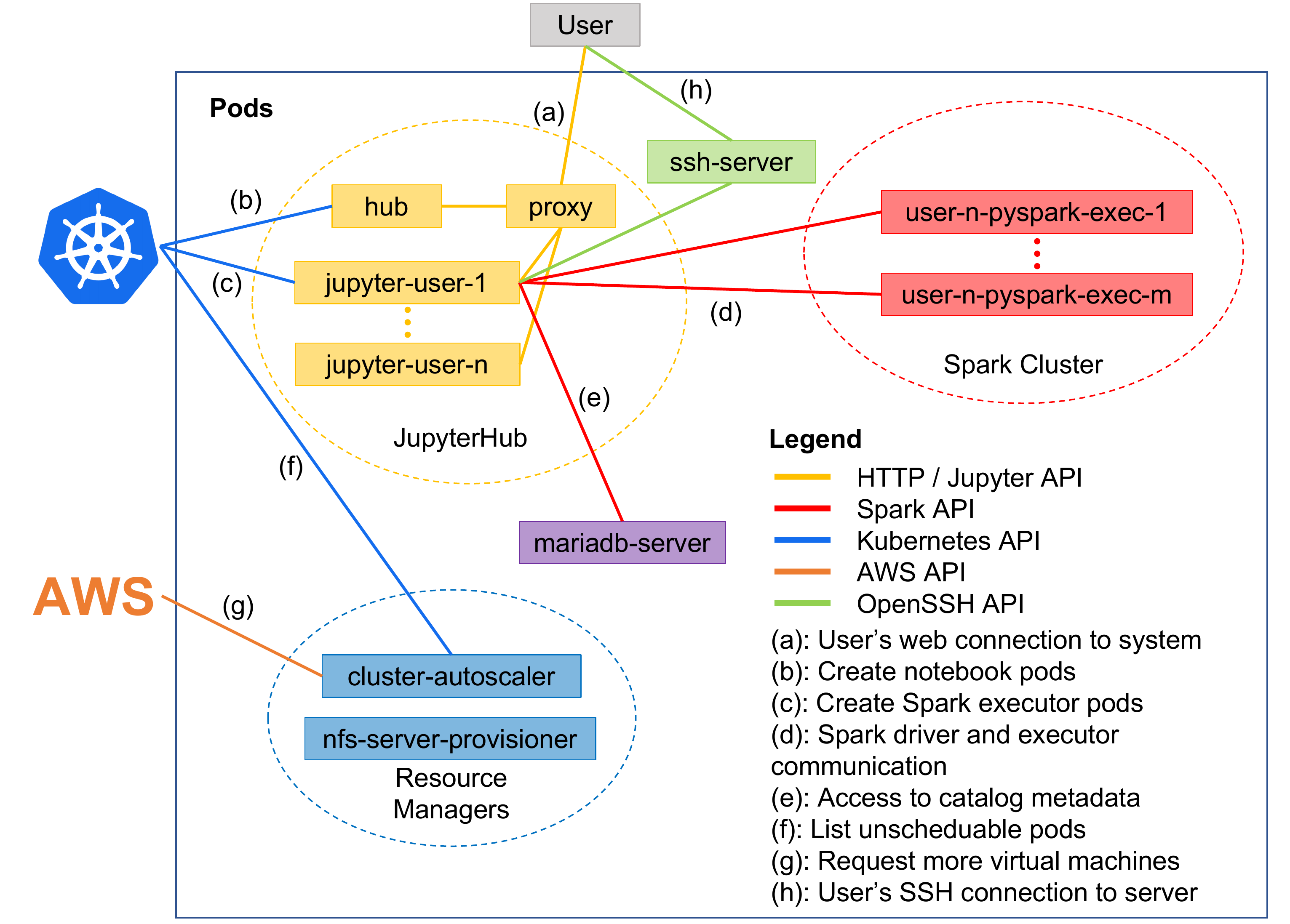}}
\caption{A diagram of the essential components of the Kubernetes cluster when the science platform is in use. Each box represents a single Kubernetes Pod scheduled on the cluster. The colors of the boxes and the dashed ovals surrounding the three groups are for visualization purposes only; each Pod exists as an independent entity to be scheduled on any available machines. The colored paths and letter markers indicate the pattern of API interactions that occur when users interact with the system. (a) shows a user connecting to the JupyterHub from the internet. The JupyterHub creates a notebook server (\texttt{jupyter-user-1}) for the user (b). The user creates a Spark cluster using their notebook server as the location for the Spark driver process (c). Scheduled Spark executor Pods connect back to the Spark driver process running in the notebook server (d). The Spark driver process accesses a MariaDB server for catalog metadata (e). In the background, the Kubernetes cluster autoscaler keeps track of the scheduling status of all Pods (f). At any point in (a)-(d), if a Pod cannot be scheduled due to a lack of cluster resources, the cluster autoscaler will request more machines from AWS to meet that need (g). Optionally, the user can connect to their running server with SSH (h).}
\label{fig:cluster}
\end{figure}

Figure~\ref{fig:cluster} shows the state of the Kubernetes cluster during normal usage of a platform created with our Helm chart as well as the pathway of API interactions that occur as a user interacts with the system. A user gains access to the system through a JupyterHub, which is a log-in portal and proxy to one or more managed Jupyter notebook servers spawned by the JupyterHub. This notebook server is run on a node of the Kubernetes cluster, which can be constrained by hardware requirements and/or administrator provided node labels. A proxy forwards external authenticated requests from the internet to a user's notebook server. Users can use the Apache Spark software, which is pre-installed on their server, to create a Spark cluster using the Spark on Kubernetes API. The user can also access their running notebook server using a Secure Shell (SSH) client.

\subsubsection{An Interface to Computing}
\label{sec:interface}

The Jupyter notebook application, and its extension Jupyter lab, provide an ideal environment for astronomers to access, manipulate, and visualize data sets. The Jupyter notebook/lab applications, although usually run locally on a user's machine, can run on a remote machine and be accessed through a JupyterHub, a web application that securely forwards authenticated requests directed at a central URL to a running notebook server.\footnote{As an example, one may access a JupyterHub at the URL \nolinkurl{https://hub.example.com} which, if you are an authenticated user, will forward through a proxy to \nolinkurl{https://hub.example.com/user/username}. When running a notebook on a local machine, there is no access to a JupyterHub and the single user server is served at (typically) \nolinkurl{http://localhost:8888}.} The authentication layer of JupyterHub allows us to block non-authenticated users from the platform. Our science platform integrates authentication through GitHub, allowing us to authenticate both individual users by their GitHub usernames and groups of users through GitHub Organization membership. For example, the implementation of this science platform described in Section \ref{sec:ztf} restricts access to the platform and its private data to members of the \texttt{dirac-institute} and \texttt{ZwickyTransientFacility} GitHub organizations.

Users can choose to bypass the Jupyter computing environment by accessing their running notebook server with a SSH client. The SSH client also facilitates file transfers between the user and their notebook server when using utilities such as \texttt{scp} or \texttt{rsync}. Access through SSH is implemented using a ``jump host'' setup: a single, always-running container on the cluster runs an OpenSSH server that is networked to the internet. When the user's notebook server is running, it additionally runs an OpenSSH server in the background. The user adds a cryptographic public key to a file system shared between the notebook server and jump host. The user connects to the jump host with their username and a cryptographic private key stored on their local machine. From the jump host, the user can connect to their running notebook server with no additional configuration. A properly formatted invocation of the \texttt{ssh} command can do this in one step. Additional public and private keys are generated automatically for each host and each user and placed on a shared file system with correct permissions.

Finally, a Virtual Network Computing (VNC) desktop is made available to the user for using graphical applications outside of the Jupyter notebook. The VNC desktop is provided through the Jupyter Remote Desktop Proxy software, an extension to the Jupyter notebook server. The in-browser desktop emulation offers reasonable interaction latencies over a typical internet connection.

\subsubsection{A Scalable Analytics Engine}
\label{sec:scalable_analytics}

Apache Spark (Spark) is a tool for general distributed computing, with a focus on querying and transforming large amounts of data, that works well in a shared-nothing, distributed computing environment. Spark uses a driver/executor model for executing queries. The driver process splits a given query into several (1 to thousands) independent tasks which are distributed to independent executor processes. The driver process keeps track of the state of the query, maintains communication with its executors, and coalesces the results of finished tasks. Since the driver and executor(s) only need to communicate with each other over the network, executor processes can remain on the same machine as a driver, to take advantage of parallelism on a single machine, or be distributed across several other machines in a distributed computing context.\footnote{Creating executor processes on a single machine isn't done in practice; instead, Spark supports multithreading in the driver process that replace the external executor process(es) when using local resources.} The API for data transformation, queries, and analysis remains the same whether or not the Spark engine executes the code sequentially on a local machine or in parallel on distributed machines, allowing code that works on a laptop to naturally scale to a cluster of computers.

To support astronomy-specific operations, \cite{zecevic} have developed the Astronomy eXtensions to Spark (AXS), a set of additional Python bindings to the Spark API to ease astronomy-specific data queries such as cross matches and sky maps in addition to an internal optimization for speeding up catalog cross matches using the ZONES algorithm, described in \cite{zecevic}. AXS is included in our science platform to ease the use of Spark for astronomers and also provide fast cross-matching capability between catalogs.

AXS requires that tabular data is stored Apache Parquet format, a compressed column-oriented data storage format.\footnote{See \url{https://parquet.apache.org/}} The columnar nature and partitioning of the files in Parquet format allows for very fast reads of large tables. For example, one can obtain a subset of just the ``RA'' column of a catalog without scanning through all parts of all of the files. Apache Spark's flexible functionality for accessing data of different formats, exposed in Python through \texttt{pyspark.sql.DataFrame.read}, allows one to convert a broad range of catalogs in different formats -- including FITS \citep{spark_fits} -- to Parquet. AXS additionally requires that catalogs stored in Parquet be similarly partitioned in order to perform fast cross-matches. AXS provides a single function, exposed in Python as \texttt{AxsCatalog.save}, that will re-partition a data frame read using Spark, save it in Parquet format, and make the table available to a user through its Apache Hive metastore database.\footnote{See \url{https://hive.apache.org/}}

\subsubsection{A Scalable Storage Solution}
\label{sec:scalable_storage}

Amazon S3 is a scalable object store with built-in backups and optional replication across geographically distinct AWS regions. Files are placed into a S3 bucket, a flat file system that scales well to simultaneous access from thousands of individual clients.  Files are accessed over the network using a REST API over HTTP, supporting actions to retrieve and create new objects in the bucket. The semantics of the S3 API are not compliant with the POSIX specification that typical file systems adhere to. However, there are projects, such as \texttt{s3fs}, that allow for mounting of the S3 object store as a traditional file system and provide an interface layer that makes the file system largely POSIX compliant.\footnote{See \url{https://github.com/s3fs-fuse/s3fs-fuse}} The names of S3 buckets are globally unique, which makes public and private sharing of data in a bucket easy: a user anywhere in the world can access public data from an S3 bucket by specifying only its name. To access private data, the user must additionally authenticate them self with AWS. Access control lists provide object-level permissions for read/write access to certain users and the public. Additionally, there is no limit to the amount of data that can be stored, although individual files must be no larger than 5 TB, and individual upload actions cannot exceed 5 GB. In this platform, we store and access TB+ tabular datasets stored in Parquet format with a common partitioning scheme, making the data AXS compatible.

\subsubsection{A deployment solution}
\label{sec:deployment}

We have created a deployment solution for organized creation and management of each of these three components. The code for this is stored at a GitHub repository accessible at \url{https://github.com/astronomy-commons/science-platform}. Files referenced in the following code snippets assume access at the root level of this repository.

To create and manage our Kubernetes cluster, we use the \texttt{eksctl} software.\footnote{See \url{https://eksctl.io/}} This software defines configuration of the Amazon Elastic Kubernetes Service (EKS) from YAML-formatted files. An EKS cluster consists of a managed Kubernetes master node that runs the control plane software along with a set of either managed or unmanaged nodegroups backed by Amazon Elastic Compute Cloud (EC2) virtual machines which the applications scheduled on the cluster.\footnote{Managed nodes are EC2 virtual machines with a tighter coupling to an EKS cluster. Unmanaged nodes allow for more configuration by an administrator.}

To help us manage large numbers of Kubernetes objects, we use Helm, the ``package manager for Kubernetes.'' Helm allows Kubernetes objects described as YAML files to be templated using a small number of parameters or ``values,'' also stored in YAML. Helm packages together YAML template files and their default template values in Helm ``charts.'' Helm charts can have versioned dependencies on other Helm charts to compose larger charts from smaller ones.  After cluster creation, we use Helm to install the \texttt{cluster-autoscaler-chart}, which deploys the Kubernetes Cluster Autoscaler application. The cluster autoscaler scales the number of nodes in the Kubernetes cluster up or down when resources are too constrained or underutilized.

We have created a Helm chart to manage and distribute versioned deployments of our  platform. This chart depends on three sub-charts: 
\begin{enumerate}
    \item The \texttt{zero-to-jupyterhub} chart, a standard and customizable installation of JupyterHub on Kubernetes. The \texttt{zero-to-jupyterhub} chart uses Docker images from the Jupyter Docker Stacks by default and uses the \texttt{KubeSpawner} for creating Jupyter notebook servers using the Kubernetes API directly.\footnote{See \url{https://jupyter-docker-stacks.readthedocs.io/} and \url{https://jupyterhub-kubespawner.readthedocs.io/}}
    \item The \texttt{nfs-ganesha-server-and-external-provisioner} chart, which provides a network filesystem server and Kubernetes-compliant storage provisioner.\footnote{See \url{https://github.com/kubernetes-sigs/nfs-ganesha-server-and-external-provisioner}}
    \item A \texttt{mariadb} chart, which provides a MariaDB server and is used as an Apache Hive metadata store for AXS.\footnote{See \url{https://mariadb.org/} and \url{https://github.com/bitnami/charts/tree/master/bitnami/mariadb}}
\end{enumerate}
The Helm chart contains configuration of the three sub-charts. For example, the chart is configured to use a Docker image with installations of Spark/AXS, the OpenSSH client/server, and Jupyter notebook server extensions like the Jupyter Remote Desktop Proxy when the JupyterHub starts a notebook server. Additional configuration in the chart provides instructions for mounting the network file system, instructions for setting up the Hive metastore, defines reasonable defaults for using Spark on Kubernetes, and defines notebook server startup-scripts that start the SSH server and set up the user's space on the file system (such as copying example notebooks to the user's home directory).

\subsection{Providing a shared filesystem with granular access control}
\label{sec:nfs}

We found it to be critically important to provide a way for users to easily share files with one another. The default Helm chart and \texttt{KubeSpawner} configuration creates a Persistent Volume Claim backed by the default storage device configured for the Kubernetes cluster for each single user server, allowing a user's files to persist beyond the lifetime of their server. For AWS, the default storage device is an EBS volume, roughly equivalent to a network-connected SSD with guaranteed input/output capabilities. By default, this volume is mounted at the file system location \texttt{/home/jovyan} in the single user container. This setup makes it difficult for the users' results to be shared with others because: a) they are isolated to their own disk, and b) by default all users share the same username and IDs, making granular access control extremely difficult.

\begin{figure}
    \centering
    \includegraphics[width=\linewidth]{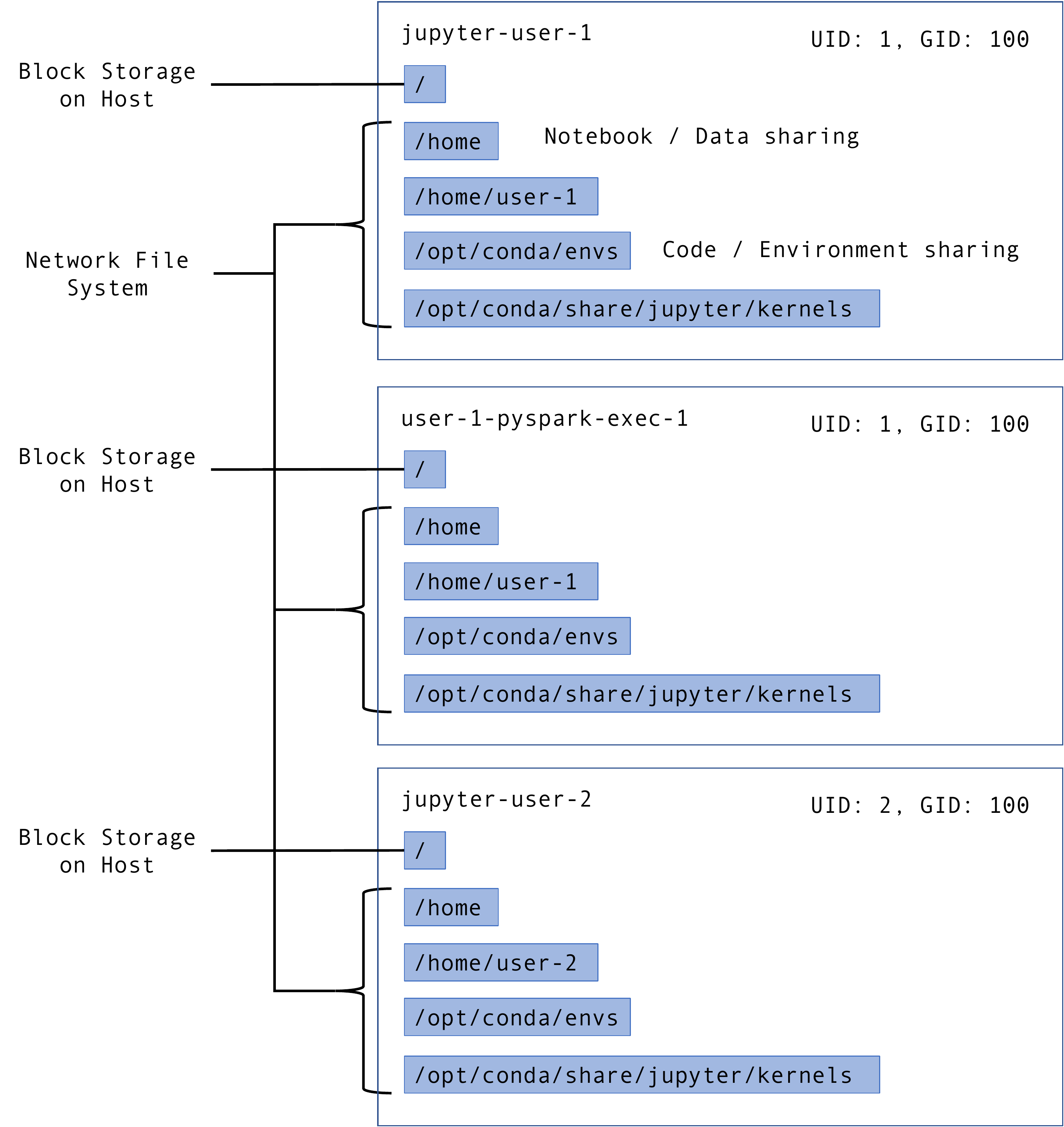}
    \caption{An illustration of the filesystem within each  container spawned by the JupyterHub (\texttt{jupyter-user-1} and \texttt{jupyter-user-2}) and by the user in the creation of a distributed Spark cluster. Most of the filesystem (the root directory \texttt{/}) exists on an ephemeral storage device tied to the host machine. The home directories, \texttt{conda} environment directories, and Jupyter kernel directories within each container are mounted from an external NFS server. This file structure allows for sharing of Jupyter Notebook files and code environments with other users and with a user's individual Spark Cluster. UNIX user ids (\texttt{UID}) and group ids (\texttt{GID}) are set to prevent unauthorized data access and edits.
    }
    \label{fig:nfs_mounts}
\end{figure}

To resolve these issues, we provisioned a network file system (NFSv4) server using the \texttt{nfs-ganesha-server-and-external-provisioner} Helm chart, creating a centralized location for user files and enabling file sharing between users. To solve the problem of access control, each notebook container is started with two environment variables: \texttt{NB\_USER} set equal to the user's GitHub username, and \texttt{NB\_UID} set equal to the user's GitHub user id. The start-up scripts included in the default Jupyter notebook Docker image use the values of these environment variables to create a new Linux user, move the home directory location, update home directory ownership, and update home directory permissions from their default values. Figure~\ref{fig:nfs_mounts} shows how the NFS server is mounted into single user pods to enable file sharing. The NFS server is mounted at the \texttt{/home} directory on the single user server, and a directory is created for the user at the location \texttt{/home/<username>}. Each user's directory is protected using UNIX-level file permissions that prevent other users from making unauthorized edits to their files. System administrators can elevate their own permissions (and access the back-end infrastructure arbitrarily) to edit user files at will. The UNIX user ids (UIDs) are globally unique, since they are equal to a unique GitHub ID.

In initial experiments, we used the managed AWS Elastic File System (EFS) service to enable file sharing. Using the managed service provides significant benefits, including unlimited storage, scalable access, and automatic back-ups. However, EFS had a noticeable latency increase per Input/Ouput operation compared to the EBS-backed storage of the Kubernetes-managed NFS server. In addition, EFS storage is $3\times$ more expensive than EBS storage.\footnote{The cost of EFS is \$0.30/GB-Month vs \$0.10/GB-Month for EBS. Lifecycle management policies for EFS that move infrequently used data to a higher-latency access tier can reduce costs to approximately the EBS level.}

In addition to storing home directories on the NFS server, we have an option to store all of the science analysis code (typically managed as \texttt{conda} environments) on the NFS server. This has several advantages relative to the common practice of keeping the code in Jupyter notebook Docker images. The primary advantage is that this allows for updating of installed software in real-time, and without the need to re-start user servers. A secondary advantage is that the Docker images become smaller and faster to download and start up (thus improving the user experience). The downside is decreased scalability: the NFS server includes a central point, shared by all users of the system. Analysis codes are often made up of thousands of small files, and a request for each file when starting a notebook can lead to large loads on the NFS server. This load increases when serving more than one client, and may not be a scalable beyond serving a few hundred users.

For systems requiring significant scalability, a hybrid approach of providing a base \texttt{conda} environment in the Docker image itself in addition to mounting user-created and user-managed \texttt{conda} environments and Jupyter kernels from the NFS server is warranted. This allows for fast and scalable access to the base environment while also providing the benefit of shared code bases that can be updated in-place by individual users.

\subsection{Providing Optimal and Specialized Resources}
\label{sec:resources}

Some users require additional flexibility in the hardware available to match their computing needs. To accommodate this, we have made deployments of this system that allow users to run their notebooks on machines with more CPU or RAM or with specialty hardware like Graphics Processing Units (GPUs) as they require. This functionality is restricted to deployments where we trust the discretion of the users and is not included in the demonstration deployment accompanying this manuscript.

\begin{figure}
    \centering
    \includegraphics[width=\linewidth]{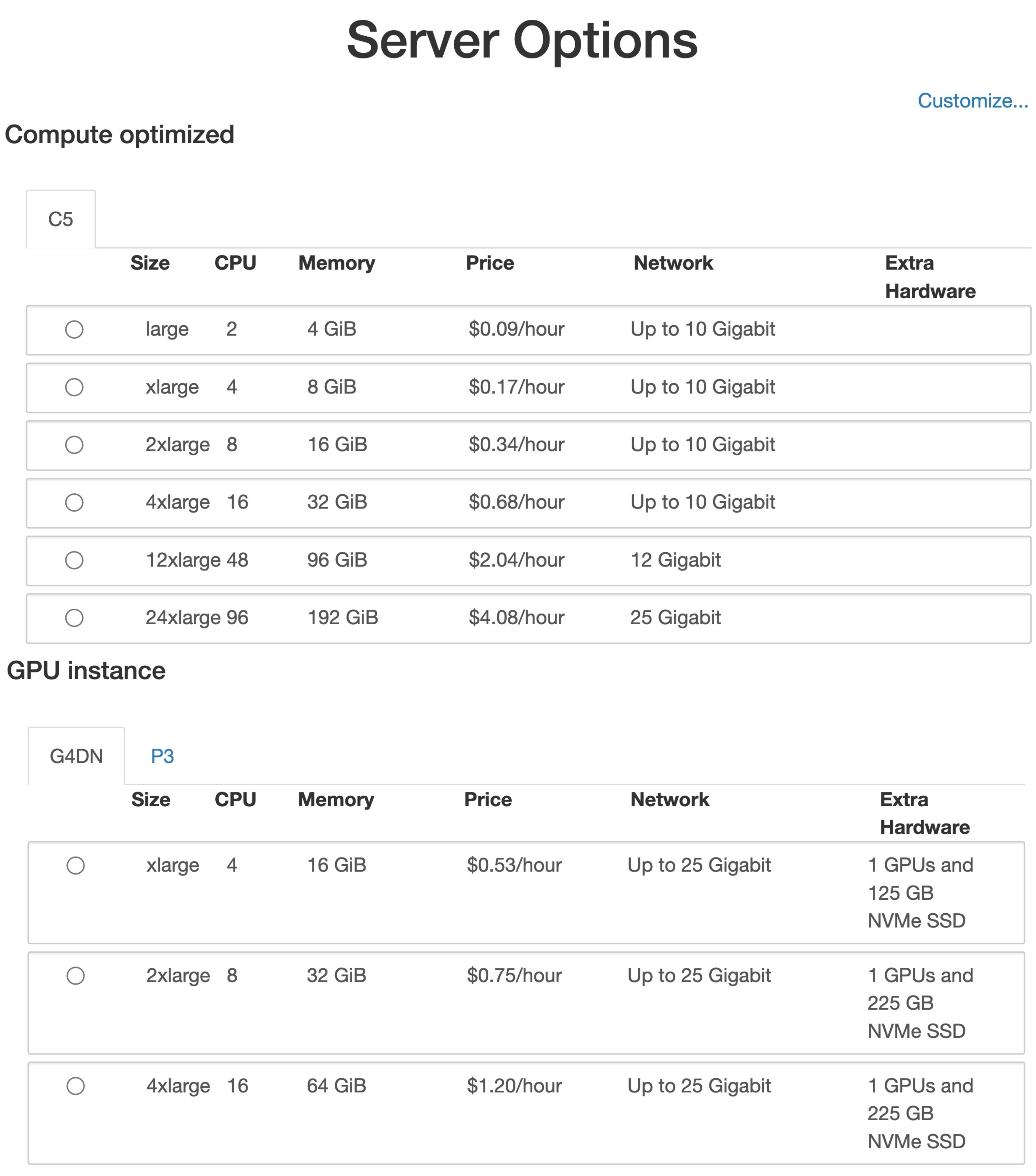}
    \caption{A screenshot of the JupyterHub server spawn page. Several options for computing hardware are presented to the user with their hardware and costs enumerated. Of note is the ability to spawn GPU instances on demand. When a user selects one of these options, their spawned Kubernetes Pod is tagged so that it can only be scheduled on a node with the desired hardware. If a node with the required hardware does not exist in the Kubernetes cluster, the cluster autoscaler will provision it from the cloud provider (introducing a ${\sim}5$ minute spawn time).}
    \label{fig:spawn_page}
\end{figure}

Flexibility in hardware is provided through a custom JupyterHub options form that is shown to the user when they try to start their server. An example form is shown in Fig.~\ref{fig:spawn_page}. Several categories of AWS EC2 instances are enumerated with their hardware and costs listed. Hardware is provisioned in terms of vCPU, or ``virtual CPU,'' roughly equivalent to one thread on a hyperthreaded CPU. In this example, users can pick an instance that has as few resources as 2 vCPU and 1 GiB of memory at the lowest cost of \$0.01/hour (the \texttt{t3.micro} EC2 instance), to a large-memory machine with 96 vCPU and 768 GiB of memory at a much larger cost of \$6.05/hour (the \texttt{r5.24xlarge} EC2 instance). In addition, nodes with GPU hardware are provided as an option at moderate cost (4 vCPU, 16 GiB memory,  1 NVIDIA Tesla P4 GPU at \$0.53/hour; the \texttt{g4dn.xlarge} EC2 instance). These GPU nodes can be used to accelerate code in certain applications such as image processing and machine learning. For this deployment, the form is configured to default to a modest choice with 4 vCPU and 16 GiB of memory at a cost of \$0.17/hour (the \texttt{t3.xlarge} EC2 instance). This range of hardware options and prices will change over time; the list provided is simply an example of the on-demand heterogeneity provided via AWS.

\subsection{Multi-cloud support}\label{sec:multi_cloud}

It is unlikely, and perhaps undesirable, that all scientists and organization will agree to use a single cloud provider when storing data, accessing computing resources, or deploying our system. There are many clouds outside of Amazon Web Services that scientists may have already chosen for computing and data storage based on factors such as the availability of compute credits, academic institutional tie-ins, convenience, familiarity, or differences in product offerings.\footnote{Google Cloud Platform (GCP), Microsoft Azure, IBM Cloud, DigitalOcean, and the National Science Foundation funded Jetstream Cloud is a short list of cloud providers.} Therefore, it is necessary to think about and accommodate multi-cloud support in our architecture. The system architecture outlined in \ref{sec:arch} is sensitive in two places to the choice of cloud provider: the deployment solution and the storage solution. 

The deployment solution we use is tied to the choice of cloud provider only during creation of the Kubernetes cluster with the deployment scripts. Helm interacts with the Kubernetes cluster directly and not the cloud provider, so it remains cloud-agnostic. For cloud providers that offer a managed Kubernetes service, then they typically offer command line interface (CLI) tools for creating and managing a Kubernetes cluster. In other cases, tools like \texttt{kops} and \texttt{Kubespray} enable cluster creation on a wide variety of public clouds as well on private computing clusters. Our deployment scripts can be extended in the future to accommodate other clouds using these tools.

The storage solution has a tighter coupling to the cloud provider that leads to potential lock-in with the cloud provider as well as issues with sharing data across clouds. While we have chosen Amazon S3 as our storage solution, similar object storage products from other cloud providers can be used. Apache Spark can access data stored in any object store that uses the S3 API. Some cloud providers offer object storage products that expose APIs that are compliant or complementary to the S3 API, which makes both accessing data between clouds feasible. For example, both Google Cloud Platform (GCP) and DigitalOcean provide storage services that have some or full interoperability with S3. This means a user can expect to deploy our system on a cloud with a compatible object store and use that object store with few or no changes in how the data are accessed. However, transferring large amounts of data between clouds through the internet (referred to as ``egress'') remains very costly, making multi-cloud data access infeasible in practice. AWS quotes data transfer fees of \$0.05-\$0.09 per GB depending on the total volume transferred in a month. This means that a user who has deployed our system in a cloud other than AWS cannot expect to access large amounts of data stored within Amazon S3. This sets the expectation that our system will be deployed in the cloud where a user's data and potentially other relevant and desirable data sets are located. Notably, this challenge persists, at a smaller scale, within an individual cloud due to data transfer costs between geographically distinct data centers (often called regions). AWS quotes data transfer costs between regions at \$0.01-\$0.02 per GB. However, unless very low latency for data access from many countries/continents is required, a user or organization can likely choose and stick to a single region when storing data and acquire computing resources.

Slightly different system architectures allow for easier multi-cloud data access. For example, the Jupyter Kernel Gateway and Jupyter Enterprise Gateway projects can be used to access computing resources that are distributed across multiple clusters.\footnote{See \url{https://github.com/jupyter-server/kernel_gateway} and \url{https://github.com/jupyter-server/enterprise_gateway}} Both of these projects provide a method to create and access a running process in a remote cluster. This allows one to create several Kubernetes clusters in different clouds where desirable datasets are located and use a single JupyterHub as an entrypoint to access data stored in multiple clouds. While this proposed solution does not bring the data closer together, which would be desirable for applications that require jointly analyzing datasets from multiple sources in different clouds, it does allow for baseline multi-cloud data access. True multi-cloud data access is likely to remain infeasible without significant decreases in egress costs or prior agreement on where to store data from dataset stakeholders. New services, such as Cloudflare R2, that provide cloud storage with zero or near-zero egress cost brighten the prospects for cheap, multi-cloud data transfer and would lift the requirement for consensus among stakeholders.

\section{A Deployment for ZTF Analyses}
\label{sec:ztf}

To demonstrate the capabilities of our system and verify its utility to a science user, we deployed to enable the analysis of data from the Zwicky Transient Facility (ZTF). Section \ref{sec:ztf_data} describes the catalogs available through this deployment, Section \ref{sec:ztf_access} demonstrates the typical access pattern to the data using the AXS API, and Section \ref{sec:ztf_science} showcases a science project executed on this platform.

\subsection{Catalogs available}
\label{sec:ztf_data}

Table~\ref{tab:datasets} enumerates the catalogs available to the user in this example deployment. We provide a catalog of light curves from ZTF, created from de-duplicated match files. The most recent version of these match files have a data volume of ${\sim}$ 4 TB describing light curves of ${\sim}$ 1 billion+ objects in the ``g'', ``r'', and ``i'' bands. In addition, we provide access to catalogs from the data releases of the SDSS, Gaia, AllWISE, and Pan-STARRS surveys for convenient cross matching. The system allows users to upload, cross match, and share custom catalogs in addition to the ones provided, using the method described in \ref{sec:scalable_analytics}.

\begin{table}
    \centering
    \begin{tabular}{c|c|c}
        \textbf{Name} & \textbf{Data Size (GB)} & \textbf{\# Objects ($10^9$)}  \\\hline\hline
        SDSS & 65 & 0.77 \\
        AllWISE & 349 & 0.81 \\
        Pan-STARRS 1 & 402 & 2.2 \\
        Gaia DR2 & 421 & 1.8 \\
        ZTF & 4100 & 1.2 \\
        \hline\hline
        Total & 5337 & 8.9
    \end{tabular}
    \caption{The sizes of each of the catalogs available on the ZTF science platform along with the total data volume.}
    \label{tab:datasets}
\end{table}

\subsection{Typical workflow}
\label{sec:ztf_access}

Users can query the available catalogs through the AXS/Spark Python API. For example, a user loads a reference to the ZTF catalog like so:
\begin{lstlisting}[gobble=0]
import axs
from pyspark.sql import SparkSession
spark = SparkSession.builder.getOrCreate()
catalog = axs.AxsCatalog(spark)
ztf = catalog.load('ztf')
\end{lstlisting}
The \texttt{spark} object represents a Spark SQL Session and a connection to a Hive metastore database, which stores metadata for accessing the catalogs. This object is used as a SQL backend when creating the \texttt{AxsCatalog}, which acts as an interface to the available catalogs. Catalogs from the metastore database are loaded by name using the AXS API. Data subsets can be created by selecting one or more columns:
\begin{lstlisting}[gobble=0]
ztf_subset = ztf.select('ra', 'dec', 'mag_r')
\end{lstlisting}
\texttt{AxsCatalog} Python objects can be crossmatched with one another to produce a new catalog with the crossmatch result:
\begin{lstlisting}[gobble=0]
gaia = catalog.load('gaia')
xmatch = ztf.crossmatch(gaia)
\end{lstlisting}
The \texttt{xmatch} object can be queried like any other \texttt{AxsCatalog} object. Spark allows for the creation of User-Defined Functions (UDFs) that can be mapped onto rows of a Spark DataFrame. The following example shows how a Python function that converts an AB magnitude to its corresponding flux in janskys can be mapped onto all ${\sim}$63 billion r-band magnitude measurements from ${\sim}$1 billion light curves in the ZTF catalog (in parallel):
\begin{lstlisting}[gobble=0]
from pyspark.sql.functions import udf
from pyspark.sql.types import ArrayType
from pyspark.sql.types import FloatType
import numpy as np

@udf(returnType=ArrayType(FloatType()))
def abMagToFlux(m):
    flux = ((8.90 - np.array(m))/2.5)**10
    return flux.tolist()
ztf_flux_r = ztf.select(
    abMagToFlux(ztf['mag_r']).alias("flux_r")
)
\end{lstlisting}

\subsection{Science case: Searching for Boyajian star Analogues}
\label{sec:ztf_science}
We test the ability of this platform to enable large-scale analysis by using it to search for Boyajian star \citep{boyajian} analogs in the ZTF catalog. The Boyajian star, discovered with the Kepler telescope, dips in its brightness in an unusual way. We intend to search the ZTF catalog for Boyajian-analogs, other stars that have anomalous dimming events, which will be fully described in Boone et al.~(in prep.); here we limit ourselves to aspects necessary for the validation of the analysis system. The main method for our Boyajian-analog searches relies on querying and filtering large volumes of ZTF light curves using AXS and Apache Spark in search of the dimming events. Objects of interest are then spatially cross matched against the other catalogs available, for example to the Gaia catalog to create a color-magnitude diagram and the AllWISE catalog to identify if there is excess flux in the infrared. This presents an ideal science-case for our platform: the \textit{entire ZTF catalog} must be queried, filtered, analyzed, and compared to other catalogs repeatedly in order to complete the science goals. 

\begin{figure}
    \centering
    \includegraphics[width=\linewidth]{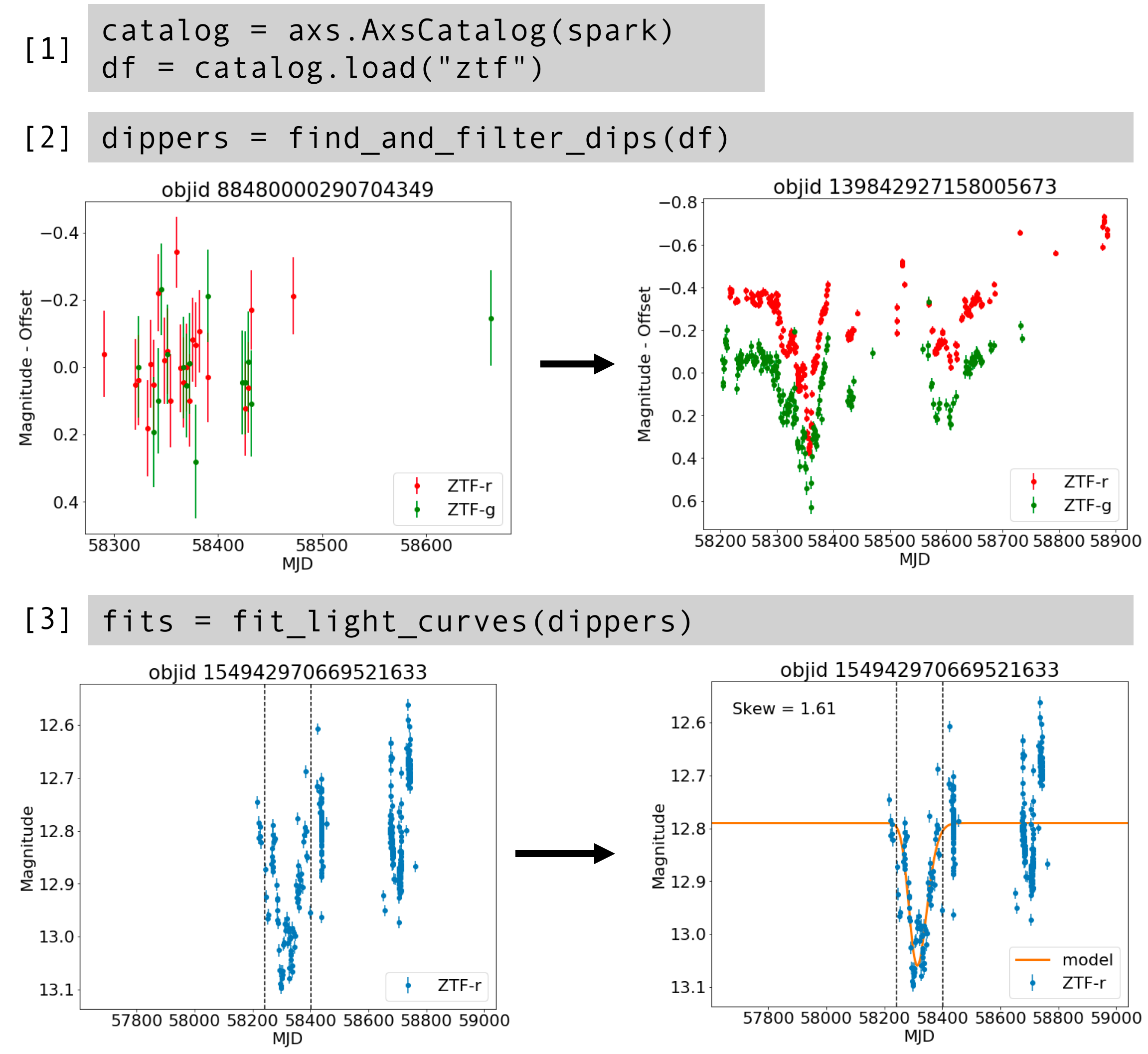}
    \caption{An example analysis (boiled down to two lines) that finds light curves in the ZTF light curve catalog with a dimming event. (1) shows how the ZTF catalog is loaded as a Spark DataFrame (\texttt{df}), (2) shows the product of filtering light curves for dimming events, and (3) shows the result of fitting a model to the remaining light curves. This process exemplifies that analyses can often be represented as a filtering and transformation of a larger dataset, a process that Spark can easily execute in parallel.}
    \label{fig:dippers}
\end{figure}

We wrote custom Spark queries that search the ZTF catalog for dimming events. After filtering the light curves, we created a set of UDFs for model fitting that wrap the optimization library from the \texttt{scipy} package. These UDFs are applied to the filtered light curves to parallelize least-squared fitting routines of various models to the dipping events. Figure~\ref{fig:dippers} shows an outline of this science process using AXS.

The use of Apache Spark speeds up queries, filtering, and fitting of the data tremendously when deployed in a distributed environment. We used a Jupyter notebook on our platform to allocate a Spark cluster of consisting of 96 \texttt{t3.2xlarge} EC2 instances. Each instance had access to 8 threads running on an Intel Xeon Platinum 8000 series processor with 32 GiB of RAM, creating a cluster with 768 threads and 3,072 GiB of RAM. We used the Spark cluster to complete a complex filtering task on the full 4 TB ZTF data volume in ${\sim}$three hours. The underlying system was able to scale to full capacity within minutes, and scale down once the demanding query was completed just as fast, providing extreme levels of parallelism at minimal cost. The total cost over the time of the query was ${\sim}$\$100.

This same complex query was previously performed on a large shared-memory machine at the University of Washington with two AMD EPYC 7401 processors and 1,024 GiB of RAM. The query utilized 40 threads and accessed the catalog from directly connected SSDs. This query previously took a full two days to execute on this hardware in comparison to the ${\sim}$three hours on the cloud based science platform. Performing an analysis of this scale would not be feasible if performed on a user's laptop using data queried over the internet from the ZTF archive.

In addition, the group was able to gain the extreme parallelism afforded by Spark without investing a significant amount of time writing Spark-specific code. The majority of coding time was spent developing science-motivated code/logic to detect, describe, and model dipping events within familiar Python UDFs and using familiar Python libraries. In alternative systems that provide similar levels of parallelism, such  as HPC systems based on batch scheduling, a user would typically have to spend significant time altering their science code to conform with the underlying software and hardware that enables their code to scale. For example, they may spend significant time re-writing their code in a way that can be submitted to a batch scheduler like PBS/Slurm, or spend time developing a leader/follower execution model using a distributed computing/communication framework such as OpenMPI. Traditional batch scheduling systems running on shared HPC resources typically have a queue that a user's program must wait in before execution. In contrast, our platform scales on-demand to the needs of each individual user.

This example demonstrates the utility of using cloud computing environments for science: when science is performed on a platform that provides on-demand scaling using tools that can distribute science workloads in a user-friendly manner, time to science is minimized.

\section{Scalability, Reliability, Costs, and User Experience}

Our system is expected to scale both in the number of simultaneous users and to the demands of a single user's analysis. In the former case, JupyterHub and its built in proxy can scale to access by hundreds of users as its workload is limited to routing simple HTTP requests. In the latter case, data queries by individual users are expected to scale to very many machines, allowing for fast querying and transformation of very large datasets. Section \ref{sec:scaling} summarizes tests to verify this claim.

\subsection{Scaling Performance}
\label{sec:scaling}

We performed scaling tests to understand and quantify the performance of our system. We tested both the ``strong scaling'' and ``weak scaling'' aspects of a simple query. Strong scaling indicates how well a query with a fixed data size can be sped up by increasing the number of cores allocated to it. On the other hand, weak scaling indicates how well the query can scale to larger data sizes; it answers the question ``can I process twice as much data in the same amount of time if I have twice as many cores?'' 

\begin{figure*}
    \centering
    \includegraphics[width=\linewidth]{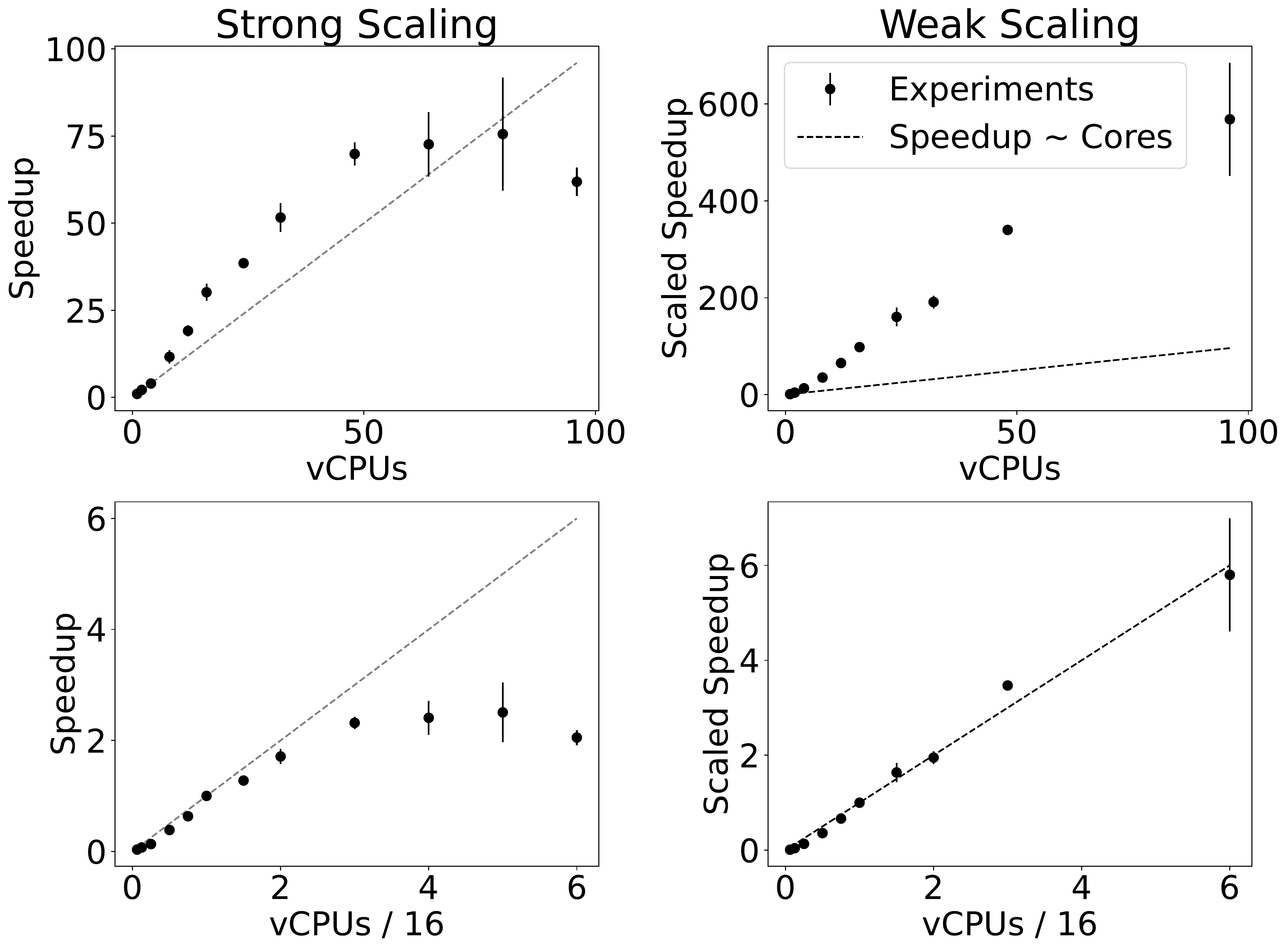}
    \caption{Speedup computed in strong scaling (left) and weak scaling (right) experiments of a simple Spark query that summed a single column of the ZTF catalog, ${\sim} 3\times10^9$ rows. Speedup is computed using Eq.~\ref{eqn:speedup} and scaled speedup is computed using Eq.~\ref{eqn:scaled_speedup}. For each value of vCPU, the query was executed several (3+) times. For each trial, the runtime  was measured and speedup calculated. Each point represents the mean value of speedup and error bars indicate the standard deviation. The first row shows speedup computed using sequential computing (vCPU $= 1$) to set the reference time and reference problem size. The second row shows speedup computed using $16$ vCPU to set the reference. With sequential computing as the reference, we observe speedup that is abnormally high in both the strong and weak scaling case. By adjusting the reference point to vCPU $= 16$, we find that we can recover reasonable weak scaling results and expected strong scaling results for a small to medium number of cores. Using the adjusted reference, we observe in the strong scaling case diminishing returns in the speedup as the number of cores allocated to the query increases, as expected. The weak scaling shows optimistic results; the speedup scales linearly with the catalog size as expected.}
    \label{fig:scaling}
\end{figure*}

Figure~\ref{fig:scaling} shows the strong and weak scaling of a simple query, the sum of the ``RA'' column of a ZTF light curve catalog, which contains ${\sim}3\times10^9$ rows, stored in Amazon S3. This catalog is described in more detail in section \ref{sec:ztf_data}. In these experiments, speedup is computed as
\begin{equation}
    \text{speedup} = t_{\text{ref}} / t_{N}
    \label{eqn:speedup}
\end{equation}
where $t_{\text{ref}}$ is the time taken to execute the query with a reference number of cores while $t_{N}$ is the time taken with $N$ cores. For the weak scaling tests, scaled speedup is computed as
\begin{equation}
    \text{scaled speedup} = t_{\text{ref}} / t_{N} \times P_{N} / P_{\text{ref}}
    \label{eqn:scaled_speedup}
\end{equation}
which is scaled by the problem size $P_N$ with respect to the reference problem size $P_{\text{ref}}$. We chose to scale the problem size directly with the number of cores allocated; the $96$-core query had to scan the entire catalog, while the $1$-core query had to scan only $1/96$ of the catalog. Typically, the reference number of cores is $1$ (sequential computing), however we noticed anomalous scaling behavior at low numbers of cores, and so we set the reference to $16$ in Fig.~\ref{fig:scaling}.

In our experiments, we used \texttt{m5.large} EC2 instances to host the Spark  executor processes, which have 2 vCPU and 8 GiB of RAM allocated to them. The underlying CPU is an Intel Xeon Platinum 8000 series processor. The Spark driver process was started from a Jupyter notebook server running on a \texttt{t3.xlarge} EC2 instance with 4 vCPU and 16 GiB of RAM allocated to it. The underlying CPU is an Intel Xeon Platinum 8000 series processor. Single \texttt{m5.large} EC2 instances have a network bandwidth of 10 Gbit/s while the \texttt{t3.xlarge} instance has a network bandwidth of 1 Gbit/s. Amazon S3 can sustain a bandwidth of up to 25 Gbit/s to individual Amazon EC2 instances. Both the data in S3 and all EC2 instances lie within the same AWS region, us-west-2. The \texttt{m5.large} EC2 instances were spread across three ``availability zones'' (separate AWS data centers): us-west-2a, us-west-2b, and us-west-2c. This configuration of heterogeneous instance types, network speeds, and even separate instance locations represent a typical use-case of cloud computing and offers illuminating insight into performance of this system with these ``worst-case'' optimization steps. 

The weak scaling test showed that scaled speedup scales linearly with the number of cores provisioned for the query; twice the data can be processed in the same amount of time if using twice the number of cores. In other words, for this query, the problem of ``big data'' is solved simply by using more cores. The strong scaling test showed expected behavior up to $\text{vCPU}/16 = 5$. Speedup increased monotonically with diminishing returns as more cores were added. Speedup dropped from $2.50$ with $\text{vCPU}/16 = 5$ to $2.05$ with $\text{vCPU}/16 = 6$, indicating no speedup can be gained beyond $\text{vCPU}/16 = 5$. Drops in speedup in a strong scaling test are usually due to real world limitations of the network connecting the distributed computers. As the number of cores increases, the number of simultaneous communications and the amount of data shuffled between the single Spark driver process and the many Spark executor processes increases, potentially reaching the latency and bandwidth limits of the network connecting these computers.

\subsection{Caveats to Scalability}

As mentioned in section \ref{sec:nfs}, the use of a shared NFS can limit scalability with respect to the number of simultaneous users. We recommend the administrators of new deployments of our platform consider the access pattern of user data and code on NFS to guarantee scalability to their desired number of users. Carefully designed hybrid models of code and data storage that utilize NFS, EFS, and the Docker image itself (stored on EBS) can be developed that will likely allow the system scale to access from hundreds of users.

\subsection{Reliability}

In general, the system is reliable if individual components (i.e. virtual machines or software applications) fail. Data stored in S3 are in practice 100\% durable; at the time of writing, AWS quotes ``99.999999999\% durability of objects over a given year.'' Data stored in the EBS volume backing the NFS server are similarly durable -- 99.999\% at the time of writing. We choose to back up these data using Amazon EBS snapshots on a daily basis so we can recover the volume in the event of volume deletion or undesirable changes.

Kubernetes as a scheduling tool is resilient to failures of individual applications. Application failures are resolved by rescheduling the application on the cluster, perhaps on another node, until a success state is reached. When the Kubernetes cluster autoscaler is used, then the cluster becomes resilient to the failure of individual nodes. Pods that are terminated from a node failure will become unschedulable, which will trigger the cluster autoscaler to scale the cluster up to restore the original size of the cluster. For example, if the user's Jupyter notebook server is unexpectedly killed due to the loss of an EC2 instance, it will re-launch on another instance on the cluster, with loss of only the memory contents of the notebook server and the running state of kernels. The same is true of each of the individual JupyterHub and Spark components. Apache Spark is fault-tolerant in its design, meaning a query can continue executing if one or all of the Spark executors are lost and restarted due to loss of the underlying nodes. Similar loss of the driver process (on the Jupyter notebook server) results in the complete loss of the query.

We have run different instances of this platform for approximately three years in support of science workloads at UW, the ZTF collaboration, a number of hackathons, and for the LSST science collaborations. Over that period, we have experienced no loss of data or nodes.

\subsection{Costs}

This section enumerates the costs associated with running this specific science platform. Since cloud computing costs can be variable over time, the costs associated with this science platform are not fixed. In this section, we report costs at the time of manuscript submission as well as general information about resource usage so costs can be recomputed by the reader at a later date.

We describe resource usage along two axes: interactive usage and core hours for data queries. Interactive usage encompasses using a Jupyter notebook server for making plots, running scripts and small simulations, and collaborating with others. Data queries encompass launching a distributed Spark cluster to access and analyze data provided on S3, similarly to the methods described in Sec.~\ref{sec:ztf_science}. Equation \ref{eqn:cost} provides a formula for computing expected monthly costs given the number of users $N_u$, the cost of each user node $C_u$, the cost of the Spark cluster nodes $C_s$, the estimated time spent per week on the system $t_u$, and the number of node hours used by each user for Spark queries in a month $t_s$: 
\begin{align}
    \text{Cost}_{\text{storage}} &= N_u \times 200 \times 0.08 \times (t_u \times (30/7) + t_s) \nonumber,\\
    \text{Cost}_{\text{machines}} &= N_u \times (C_u \times t_u \times (30/7) + C_s \times t_s) \nonumber,\\
    \text{Cost} &= \text{Cost}_{\text{storage}} + \text{Cost}_{\text{machines}}. \label{eqn:cost}
\end{align}
Fixed in the equation are constants describing the amount (200 GB) and cost of ($\$0.08$/GB/month) of EBS-backed storage allocated for each virtual machines. Additionally, the term $(30/7)$ converts weekly costs to monthly costs. Node hours can be converted to core hours by multiplying $t_s$ by the number of cores per node.

Table \ref{tab:cost} enumerates the fixed costs of the system as well as the variable costs, calculated using Eq.~\ref{eqn:cost}, assuming different utilization scenarios, varying the number of users ($N_u$), the amount interactive usage per week ($t_u$), and amount of Spark query core hours each month ($t_s$). The fixed costs of the system total to $\$328.51$/month, paying for:
\begin{enumerate}
    \item a small virtual machine, \texttt{t3.medium}, for the JupyterHub web application, proxy application, and NFS server ($\$29.95$/month) with 200 GB EBS-backed storage  ($\$16.00$/month);
    \item two reserved nodes for incoming users at the default virtual machine size of \texttt{t3.xlarge} \\($\$119.81$/month) with 200 GB EBS-backed storage each ($\$32.00$/month);
    \item EBS-backed storage for the NFS server for user files ($\$8.00$/month);
    \item and storage of 5,337 GB of catalog data on Amazon S3 ($\$122.75$/month).
\end{enumerate}

\begin{table*}
    \centering
    \begin{tabular}{cccc}
        \multicolumn{4}{c}{\textbf{Virtual Machines}} \\ \hline\hline
        \textbf{Type} & \textbf{Unit Cost} & \textbf{Amount} & \textbf{Total} 
        \\ \hline
        
        Services (\texttt{t3.medium}\footnote{On-Demand pricing in region \texttt{us-west-2}: \url{https://aws.amazon.com/ec2/pricing/on-demand/}}) & \$0.0416/hour/node & 1 node & \$29.95/month
        \\ \hline
        Users (\texttt{t3.xlarge}) & \$0.1664/hour/node & 2 nodes + variable & \$119.81/month + variable
        \\ \hline
        Spark Clusters (\texttt{t3.xlarge} Spot\footnote{Spot pricing in region \texttt{us-west-2}: \url{https://aws.amazon.com/ec2/spot/pricing/}}) & \$0.0499/hour/node & variable & variable
        \\ \hline\hline \\
        
        \multicolumn{4}{c}{\textbf{Storage}} \\ \hline\hline 
        
        \textbf{Type} & \textbf{Unit Cost} & \textbf{Amount} & \textbf{Total} 
        \\ \hline
        
        Catalogs (S3\footnote{For the first 50 TB: \url{https://aws.amazon.com/s3/pricing/}}) & \$0.023/GB/month & 5,337 GB & \$122.75/month 
        \\ \hline
        
        NFS (EBS\footnote{\label{fn:ebs}General purpose SSD (gp3): \url{https://aws.amazon.com/ebs/pricing/}}) & \$0.08/GB/month & 100 GB & \$8.00/month 
        \\ \hline
        
        Node Storage (EBS) & \$0.08/GB/month/node & 200 GB/node & \$48.00/month + variable
        \\ \hline\hline \\
        
        \multicolumn{4}{c}{\textbf{Fixed Costs}}
        \\ \hline\hline
        
        \multicolumn{2}{c}{\textbf{Type}} & \multicolumn{2}{c}{\textbf{Total}}
        \\ \hline
        
        \multicolumn{2}{c}{Virtual Machines} & \multicolumn{2}{c}{\$149.76/month}
        \\ \hline
        
        \multicolumn{2}{c}{Storage} & \multicolumn{2}{c}{\$178.75/month}
        \\ \hline
        
        \multicolumn{2}{c}{All} & \multicolumn{2}{c}{\$328.51/month}
        \\ \hline\hline \\

        \multicolumn{4}{c}{\textbf{Variable Costs}}
        \\ \hline\hline
        
        \multirow{2}{*}{\textbf{Number of Users}} &  \multirow{2}{*}{\shortstack{\textbf{Interactive Usage} \\(hours/week/user)}} &  \multirow{2}{*}{\shortstack{\textbf{Spark Query Core Hours} \\ (/user/month)}} & \multirow{2}{*}{\textbf{Total}}
        \\ \\ \hline
        
        \multirow{4}{*}{10} & \multirow{2}{*}{12} & 512 
        & \$189.32/month
        \\ \cline{3-4}
        
         &  & 2048 & \$466.27/month
        \\ \cline{2-4}
        
        & \multirow{2}{*}{40} & 512 & \$415.67/month
        \\ \cline{3-4}
        
        & & 2048 & \$692.62/month
        \\ \hline
        
        \multirow{4}{*}{100} & \multirow{2}{*}{12} & 512 & \$1,893.22/month
        \\ \cline{3-4}
        
         &  & 2048 & \$4,662.71/month
        \\ \cline{2-4}
        
         & \multirow{2}{*}{40} & 512 & \$4,156.69/month
        \\ \cline{3-4}
        
         &  & 2048 & \$6,926.18/month
        \\ \hline\hline
    \end{tabular}
    \caption{
    Fixed and variable costs associated with running this analysis platform on Amazon Web Services. This summary provides cost estimates for renting virtual machine and storing data. Additional costs on the order of ${\sim} \$10$ due to network communication and data transfer are excluded from these results. Reasonable low and high estimates are chosen for the number of active users and the amount of interactive usage they have with the system. The number of Spark query core hours used by each user per month is a guess, but the high end estimate is similar to the core hours used during the analysis in Sec.~\ref{sec:ztf_science}.
    }
    \label{tab:cost}
\end{table*}

Variable costs are harder to estimate, but Table \ref{tab:cost} outlines several scenarios to get a sense for the lower/upper limits to costs. 10 scientists using the platform for 4 hours per day 3 days per 7 day week, each using 512 core hours for Spark queries each month (equivalent to 16 hours with a 32 core cluster) adds a cost of $\$189.32$/month. On the other hand, 100 scientists using the platform for 8 hours per day 5 days per 7 day week, each using 2048 core hours for Spark queries each month (64 hours with a 32 core cluster) adds a cost of $\$6,926.18$/month. There are additional costs on the order of ${\sim} \$10$ that we don't factor into this analysis. Specifically:
\begin{enumerate}
    \item network communication between virtual machines in different availability zones, introduced when scaling a Spark cluster across availability zones;
    \item data transfer costs in the form of S3 GET API requests (data transfer to EC2 virtual machines in the same region is free), introduced in each query executed against the data;
    \item and network communication between virtual machines and users over the internet, introduced with each interaction in the Jupyter notebook through the user's web browser.
\end{enumerate}
Each of these costs are minimal, and so we don't include them in our analysis. However, they are worth mentioning because they can scale to become significant. Spark queries requiring GB/TB data shuffling between driver and executors should restrict themselves to a single availability zone to avoid the costs of (1). Costs from (2) are unavoidable, but care should be taken so no S3 requests occur between different AWS regions and between AWS and the internet. Finally, (3) can balloon in size if one allows arbitrary file transfers between Jupyter servers and the user or allows large data outputs to the browser.

The number of core hours for queries is a parameter that will need to be calibrated using information about usage of this type of platform in the real-world. The upper limit guess of 2048 core hours per user per month is roughly equivalent to each user running an analysis similar to that described in Sec.~\ref{sec:ztf_science} each month. By monitoring interactive usage of our own platform and other computation tools, we estimate that realistic usage falls closer to the lower limits we provide; few users will use the platform continuously in an interactive manner, and even fewer will be frequently executing large queries using Spark.

\subsection{Dynamic Scaling}

\begin{figure*}
    \centering
    \includegraphics[width=\linewidth]{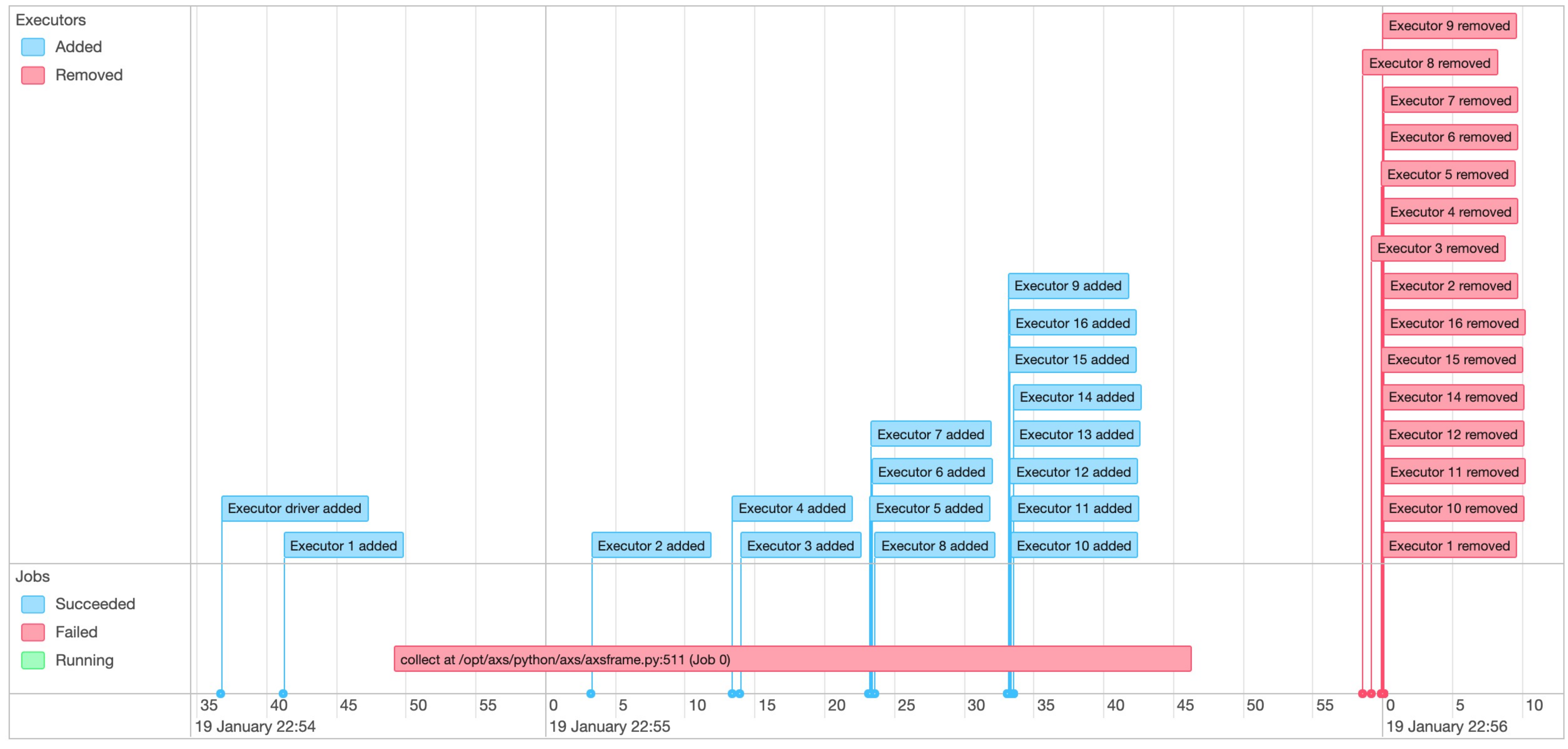}
    \caption{A screenshot of the job timeline from the Spark UI when dynamic allocation is enabled. A long-running query is started, executing with a small number of executors. As the query continues, Spark adds exponentially more executors to the cluster at a user-specified interval until the query completes or  the max number of executors is reached. Once the query completes (or is terminated, as shown here), the Spark executors are removed from the cluster.}
    \label{fig:dynamic_allocation}
\end{figure*}

Recent versions of Apache Spark provide support for ``dynamic allocation'' of Spark executors for a Spark cluster on Kubernetes.\footnote{Since Spark version 3.0.0 by utilizing shuffle file tracking on executors as an alternative to an external shuffle file service, which is awaiting support in Kubernetes. See: \url{https://spark.apache.org/docs/3.0.0/configuration.html\#dynamic-allocation}} Dynamic allocation allows for the Spark cluster to scale up its size to accommodate long-running queries as well as scale down its size when no queries are running. Figure \ref{fig:dynamic_allocation} shows pictorially this scaling process for a long-running query started by a user. This feature is expected to reduce costs associated with running Spark queries since Spark executors are added and removed based on query status, not cluster creation. This means the virtual machines hosting the Spark executor processes will be free more often either to host the Spark executors for another user's query or be removed from the Kubernetes cluster completely.

\subsection{User Experience}

While the experience of using the science platform is largely identical to using a local or remotely-hosted Jupyter notebook server, the use of containerized Jupyter notebook servers on a scalable compute resource introduces a few notable points of difference. First,  similarly to using a remotely-hosted Jupyter notebook, the filesystem exposed to the user has no direct connection to their personal computer, an experience that can be unintuitive to the user. File uploads and download can be facilitated through the Jupyter interface, but the process remains clunky. For streamlined file transfer, the user must fall back to using an SSH client and utilities like \texttt{scp} or \texttt{rsync}. In future deployments of this system, it is likely that new user interfaces will need to be produced to maximize usability of the filesystem.

Additionally, in order to allow for scale-down of the cluster, notebook servers are typically shut down after a configurable period of inactivity using the \texttt{jupyterhub-idle-culler} service. A period of $\sim1-8$ hours is typical for deployments of this science platform. This has the positive effect of reducing costs but at a detriment to the user experience. At the time of writing, inactivity is determined in terms of browser connectivity, so a user cannot expect to leave code running longer than the cull period e.g. overnight. \cite{elsa} have implemented functionality to checkpoint the memory contents of the notebook server to disk before stopping with the ability to restore the server to a running state at will. Such checkpoint/restore functionality solves the issue of interrupting running code when culling servers; however, this still does not allow for codes to run longer than the cull period. At the time of writing, additional functionality is being added to the \texttt{jupyterhub-idle-culler} service to allow for fine-grained control over which servers are culled and when.

Finally, the underlying scalable architecture introduces server start-up latencies that are noticeable to the user. Virtual machines that host notebook servers and Spark cluster executors are requested from AWS on-demand by the user. The process of requesting new virtual machines from AWS, downloading relevant Docker images to that machine, and starting the notebook/Spark Docker container can take up to ${\sim}5$ minutes.\footnote{This time is dependent on the individual cloud provider.  DigitalOcean, another cloud provider, can provision virtual machines in ${\sim}1.5$ minutes based on the experience of the authors.} The user can encounter this latency when logging onto the platform and requesting a server. They also encounter this latency when creating a distributed Spark cluster, as many machines are provisioned on-demand to run Spark executors. The log-in latency can be mitigated by keeping a small number of virtual machines in reserve so that an incoming user can instantly be assigned to a node. The \texttt{zero-to-jupyterhub} Helm chart implements this functionality through its \texttt{user-placeholder} option. This functionality schedules placeholder servers on the Kubernetes cluster that will be immediately evicted and replaced when a real user requests a server. Additionally, to avoid Docker image download times, relevant Docker images can be cached inside a custom-built virtual machine image (in AWS lingo, the Amazon Machine Image or AMI) that the virtual machine is started from. An alternative solution to this would be to place all incoming users on a shared machine, an equivalent to a ``log-in node'', before moving them to a larger machine at the user's request or automatically once a new server is provisioned from the cloud provider -- a process known as live migration. \cite{elsa} provides a path towards live migration of containerized Jupyter notebook servers, but this advanced functionality remains to be implemented with a JupyterHub deployment on Kubernetes.

\section{Conclusions} \label{sec:conclusions}

In this paper, we've described an architecture of a Cloud-based science platform as well as an implementation on Amazon Web Services that has been tested with data from the Zwicky Transient Facility. The system is shown to scale to and allow parallel analysis with $O(10\mathrm{TB})$ sized tabular, time-series heavy, datasets. It has enabled a science project that utilizes a 1 billion+ ZTF light curve catalog in full, while requiring minimal effort from domain scientists to scale their analysis from a single light curve to the full catalog. The system demonstrates the utility of elastic computing, the I/O capacity of the cloud, and distributed computing tools like Spark and AXS.

This work should be viewed in the context of exploring the feasibility of making more astronomical datasets available on cloud platforms, and providing services and platforms -- such as the one described here -- to combine and analyze them. Using this platform, it is both feasible and practical to perform large-scale cross-catalog analyses using any catalog uploaded to AWS S3 in the AXS-compatible format.\footnote{For additional practicality, catalogs must also be uploaded to the same AWS region due to constraints on data transfer costs between AWS data centers. However, given the advent of cloud storage solutions such as Cloudflare R2 with zero data transfer costs, it seems plausible that this practicality constraint may soon be lifted.} This enables any catalog provider -- whether large or small -- to make their data available to the broad community via a simple upload. Additionally, other organizations can stand up their own services on the Cloud -- either use-case specific services or broad platforms such as this-one -- to access the data using the same S3 storage API. 

In this regime, the roles of data archive and data user can be further differentiated, to the benefit of the user and perhaps at reduced cost. Data archivers upload their data to the cloud and bear the cost of storage. These costs are manageable, even by small organizations; storing 1 TB of data in S3 costs ${\sim}\$25$ per month. Cost scales dramatically when considering datasets at the PB level and timescales extending over years: a 1 PB dataset will cost ${\sim}\$3,000,000$ over 10 years assuming storage costs do not decrease. At these scales special pricing contracts may have to be negotiated between the cloud provider and archive. Additional cost scales with the number of requests for this data and the amount of data transferred. So-called ``requester-pays'' pricing models, supported by some cloud providers, can offload access and data transfer costs to the user. A user -- or perhaps an organization of users -- can deploy a system like ours at reasonable cost to access the data in a given cloud. In this case, the cost of analysis decouples from the cost of storage: it is the user who controls the number of cores utilized for the analysis, and any additional ephemeral storage used for the analysis. It is easy to imagine the user -- as a part of their grant -- being awarded cloud credits for their research, which could be applied towards these costs \citep{cloudbank}. Finally, an intermediary role may appear: the science platform provider, which has an incentive towards continuous improvements of science platforms and associated tools, which are now best viewed as systems utilized by astronomers to enable the exploration of a multitude of datasets available. The incentive of a science platform provider is to maximize science capability while minimizing the cost to the user, who now has the ability to ``shop around'' with their cloud credits for a system most responsive to their needs.

\begin{acknowledgments}
This material is based upon work supported by the National Science Foundation under Grant No. AST-2003196. A. Connolly is partially supported by NSF-1739419.

The authors acknowledge the support from the University of Washington College of Arts and Sciences, Department of Astronomy, and the DiRAC Institute. The DiRAC Institute is supported through generous gifts from the Charles and Lisa Simonyi Fund for Arts and Sciences and the Washington Research Foundation. M. Juri\'{c} wishes to acknowledge the support of the Washington Research Foundation Data Science Term Chair fund, and the University of Washington Provost’s Initiative in Data-Intensive Discovery. 

Based on observations obtained with the Samuel Oschin Telescope 48-inch and the 60-inch Telescope at the Palomar Observatory as part of the Zwicky Transient Facility project. ZTF is supported by the National Science Foundation under Grant No. AST-1440341 and a collaboration including Caltech, IPAC, the Weizmann Institute for Science, the Oskar Klein Center at Stockholm University, the University of Maryland, the University of Washington, Deutsches Elektronen-Synchrotron and Humboldt University, Los Alamos National Laboratories, the TANGO Consortium of Taiwan, the University of Wisconsin at Milwaukee, and Lawrence Berkeley National Laboratories. Operations are conducted by COO, IPAC, and UW.

The authors thank AWS Cloud Credits for Research program for supporting this project.

This material is based upon work supported by the U.S. Department of Energy, Office of Science, Office of Advanced Scientific Computing Research, Department of Energy Computational Science Graduate Fellowship under Award Number DE-SC0019323.

The authors thank the anonymous referee for their helpful review of this manuscript. S. Stetzler thanks Juan CV. Barboza, Hannah V. Bish, Thomas R. Quinn, and Jessica K. Werk for their helpful comments and assistance in preparing this manuscript.
\end{acknowledgments}

\bibliography{references}{}
\bibliographystyle{aasjournal}

\end{document}